\documentclass[twocolumn,showpacs,preprintnumbers,amsmath,amssymb]{revtex4}

\usepackage{graphicx}
\usepackage{dcolumn}
\usepackage{bm}
\usepackage{hyperref}

\newcommand{\e}{\mathrm{e}}

\renewcommand{\d}{{\rm d}}

\newcommand{\densOrd}[2]{\rho_{ #1}^{[#2]}}
\newcommand{\PhiOrd}[2]{\Phi_{ #1}^{[#2]}}

\newcommand{\Egy}{E}
\newcommand{\ksa}{k\sigma\ell}
\newcommand{\cdag}{c^\dag}
\newcommand{\cno}{c^{{}}}
\newcommand{\up}{\uparrow}
\newcommand{\down}{\downarrow}
\newcommand{\ket}[1]{|#1\rangle}
\newcommand{\bra}[1]{\langle #1|}
\newcommand{\ketbra}[2]{| #1 \rangle\!\langle #2|}

\newcommand{\ave}[1]{\left\langle #1\right\rangle}
\newcommand{\imai}{{\rm i}}

\begin{document}
\preprint{Proceedings of FQMT08, published in Physica E {\bf 42}, 595 (2010), 
\href{http://dx.doi.org/10.1016/j.physe.2009.06.069}
{doi:10.1016/j.physe.2009.06.069}}
\title{Modeling of cotunneling in quantum dot systems}
\author{Jonas Nyvold Pedersen}
\author{Andreas Wacker}
\email{Andreas.Wacker@fysik.lu.se}
\affiliation{Mathematical Physics, Lund University, Box 118, 22100 Lund, Sweden}


\begin{abstract}
Transport through nanosystems is treated within the second order
von Neumann approach. This approach bridges the gap between rate
equations which neglect level broadening and cotunneling, and the
transmission formalism, which is essentially based on the
single-particle picture thereby treating many-particle
interactions on an approximate level. Here we provide an
alternative presentation of the method in order to clarify the underlying
structure. Furthermore we apply it to the problem
of cotunneling. It is shown that both elastic and inelastic
cotunneling can be described quantitatively, while the
transmission approach with a mean-field treatment of the
interaction provides an artificial bistability.
\end{abstract}

\pacs{73.23.Hk,73.63.-b}
\keywords{transport, quantum dots, master equation}
\maketitle

\section{Introduction}
Nanostructure technology allows for the fabrication of small
structures, such as quantum dots \cite{BimbergBook1999}, nanowires
\cite{ThelanderMaterialsToday2006}, carbon nanotubes
\cite{DresselhausBook2001}, as well as molecular structures, whose
electronic properties are dominated by a small number of electrons.
The insertion into electrical circuits exhibits current-bias
($IV$) relations, which strongly differ from conventional
resistors. Typical features are the suppression of current for low
bias due to Coulomb blockade and pronounced current steps, whose
position can be easily controlled by a gate bias, thus suggesting
transistor action.

\begin{figure}[bt]
  \begin{center}
    \includegraphics[angle=0,width=.45\textwidth]{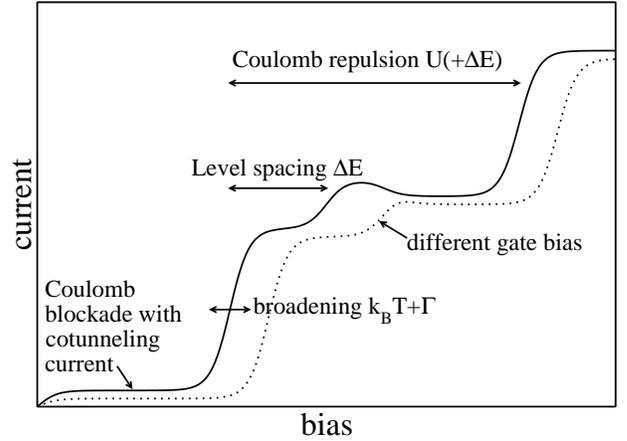}
    \caption{Schematic current-bias relation for a quantum dot, showing the
      qualitative influence of the key energy scales.}
    \label{FigOverwiew}
  \end{center}
\end{figure}

Due to the small spatial dimensions, the level quantisation with a
spacing $\Delta E$ is of fundamental relevance. Further relevant
energy scales are: The Coulomb interaction between electrons (or
similarly any other type of many-particle interaction
\cite{CapellePRL2007}) of the order $U=e^2/C$, where $C$ is the
geometrical capacitance; the tunneling rate $\Gamma/\hbar$ for the
transition of particles between the structure and the contacts; as
well as the thermal energy $k_BT$. For typical nanostructured
electronic systems studied experimentally, these energies are all
in the range of 0.1--10 meV, where $\Gamma$ can be even smaller.
Generic examples of $IV$-characteristics are shown in
Fig.~\ref{FigOverwiew}.

The task to describe the current through such structures
quantitatively is an evolved topic and a large variety of
approaches has been used within the last two decades. The general
starting point is to divide the Hamiltonian of the system as
\begin{equation}\label{EqHam}
H=H_D+H_{\text{leads}}+H_T.
\end{equation}
Here $H_D$ describes the central region (such as
a single or multiple quantum dot structure,
a molecule, a carbon nanotube, or a nanowire), which
we refer to as the quantum dot in the following.
\begin{equation}
H_{\mathrm{leads}}=\sum_{\ksa}\Egy_{\ksa}\cdag_{\ksa} \cno_{\ksa},
\end{equation}
describes the leads, where $\sigma=\up,\down$ denotes the spin,
$k$ labels the spatial wave functions of the contact states, and
$\ell$ denotes the lead (typically $\ell=L/R$ for the left and
right lead, respectively). In most works a thermal occupation
of the leads 
\[
\left\langle\cdag_{\ksa} \cno_{\ksa}\right\rangle
=f_{\ksa}=\frac{1}{\e^{(\Egy_{\ksa}-\mu_\ell)/k_BT}+1},
\]
with chemical potential $\mu_\ell$ is assumed, where the bias
$eV_\mathrm{bias}=\mu_L-\mu_R$ drives the current in a two
terminal system. Conventionally noninteracting leads are
assumed.\footnote{A finite lifetime of the lead states has
recently been addressed in \cite{KnezevicPRB2008}.} Finally, $H_T$
describes the coupling between the leads and the quantum dot.

If $H_D$ does not contain many-particle interactions, it is
straightforward to derive a transmission formula (see
\cite{DattaBook1995} and references therein) yielding a particle
current from lead $\ell$ into the quantum dot
\begin{equation}
J_\ell=\frac{1}{2\pi\hbar}\int\d E~ T_{\ell\to \ell'}(E)
\left[f_\ell(E)-f_{\ell'}(E)\right]. \label{EqTransmission}
\end{equation}
Here the transmission function is determined by the
single-particle spectrum of $H_D$ in connection with the coupling
$H_T$, which essentially broaden the levels. A formal tool is the
nonequilibrium Green function formalism \cite{HaugJauhoBook1996}.
An inclusion of many-particle effects within the mean-field level
is straightforward and exchange-correlation effects can be taken
into account by more involved mean-field potentials
\cite{BrandbygePRB2002}. Also time-dependent simulations are
possible using time-dependent density functional theory
\cite{StefanucciLNP2006}. However, hardly any results exist, which
go beyond mean-field (a rather old example is
\cite{MakoshiSurfaceScience1996}). Thus this approach works for
arbitrary temperatures if correlation effects due to the Coulomb
interaction in the system are small.

Another approach is to start from the von-Neumann equation
\begin{equation}
\imai\hbar\frac{\d}{\d t}\rho = -[\rho,H]
\label{EqvN}
\end{equation}
treating the time evolution of the density operator $\rho$ containing both the
system and the leads. As only the time-evolution of the system is of interest,
one can trace-out the lead degree of freedom in order to obtain
the reduced dot density operator
\begin{equation}
\rho_\mathrm{dot}(t)=\mathrm{Tr}_\mathrm{leads}\left\{\rho(t)\right\}.
\end{equation}
The key task is to find an approximate equation of motion for
$\rho_\mathrm{dot}(t)$ taking into account the coupling $H_T$.
This provides (generalized) master equations
\cite{BreuerBook2006}.

To lowest order in $H_T$ electron transitions between the quantum
dot and the leads are treated independently from each other (here
called first order following the number of correlated
transitions). Restricting to diagonal elements of
$\rho_\mathrm{dot}(t)$, one obtains to lowest order the Pauli
master equations \cite{BeenakkerPRB1991}.\footnote{A higher-order
expansion can be done within the diagrammatic real-time transport
theory \cite{SchoellerPRB1994}.} Furthermore, coherences,
represented by nondiagonal matrix elements, can be treated in
different ways (frequently also denoted quantum master equations).
The traditional approach is the Wangness-Bloch Redfield kinetics
\cite{WangsnessPR1953,RedfieldIBM1957} which has been applied to
quantum dots in \cite{HarbolaPRB2006,TimmPRB2008}. While
reasonable results are typically found, there is a fundamental
problem due to the occurrence of negative occupations as the
equations are not of Lindblad type \cite{LindbladCMP1976}.
Lindblad-type kinetic equations were derived in
\cite{StoofPRB1996,GurvitzPRB1996}, but only for the high-bias
limit. All first-order approaches entirely neglect broadening
effects but can treat the interactions in the system exactly by a
dia\-gonalization of the isolated dot Hamiltonian
\cite{KinaretPRB1992}
\begin{equation}
H_{D}=\sum_{a}\Egy_a|a\rangle\langle a|,
\label{EqHdot}
\end{equation}
where the states $|a\rangle$ are many-particle states, which may
be highly correlated.

In order to treat both broadening and interaction, we introduced a
second order approach based on the von Neumann equation
(\ref{EqvN}) which considers both the dynamics of
$\rho_\mathrm{dot}(t)$ and higher-order tunneling transitions
\cite{PedersenPRB2005a}. This second order von Neumann approach
(2vN)  bridges the gap between the transmission formalism, which
is fully reproduced for systems without interactions at arbitrary
temperatures, and the first order generalized master equation
schemes, which are recovered in the limits of high temperature or
large bias. 

In this paper we provide a slightly different presentation of our 2vN method
in Sec.~\ref{Sec2vN} in order to highlight its structure. 
Its range of validity and
previous results are summarized in Sec.~\ref{SecValidity}.
In Sections~\ref{SecElasCot},\ref{SecInelasCot} we show new results
demonstrating its applicability to
the cases of elastic and inelastic cotunneling, respectively.
The failure of mean-field models, providing a fictitious bistability is
addressed in Sec.~\ref{SecInelasCot} as well.

\section{The second order von Neumann approach} \label{Sec2vN}

Using $H_D$ in its diagonal representation (\ref{EqHdot}), the
tunneling between the states in the leads and the dot reads (see
Appendix~A of \cite{PedersenPRB2005a})
\begin{equation}
H_T=\sum_{\ksa,ab}T_{ba}(\ksa)\ketbra{b}{a} \cno_{\ksa}
+h.c.
\label{EqHtunnel}
\end{equation}
The matrix element $T_{ba}(\ksa)$ is the scattering amplitude for
an electron in the state $\ksa$ tunneling from the lead onto the
dot, thereby changing the dot state from the state $|a\rangle$ to
the state $|b\rangle$. Note that this amplitude vanishes unless
the number of electrons in state $|b\rangle$, $N_b$, equals
$N_a+1$. Here we use the convention that the particle number
increases with the position in the alphabet of the denoting
letter.

A general state vector for the entire system is written as $|a
g\rangle =|a\rangle\otimes|g\rangle$, with
$|g\rangle=|\{N_{k\ell\sigma}\}\rangle$ denoting the state of both
leads, where $N_{\ksa}\in \{0,1\}$. To ensure the anti-commutator
rules of the operators we use the following notation
$\ket{g-\ksa}\equiv \cno_{\ksa}\ket{g}$ and $\ket{g+\ksa}\equiv
\cdag_{\ksa}\ket{g}$. The order of indices is opposite to the
order of the operators. E.g. $\ket{g-k'\sigma'
\ell'+\ksa}=\cdag_{\ksa}\cno_{k'\sigma'\ell'}\ket{g}
=-\cno_{k'\sigma'\ell'}\cdag_{\ksa}\ket{g}
=-\ket{g+\ksa-k'\sigma'\ell'}$, taking into account the
anti-commutation rules of the operators. To simplify the notation,
$\sigma\ell$ is only attached to $k$ the first time the index $k$
appears in the equation, and in the following it is implicitly
assumed to be connected with $k$. Furthermore, we apply the
convention that $\sum_{k\sigma(\ell)}$ means summing over $k$ and
$\sigma$ with a fixed $\ell$ being
connected to $k$ in this sum.\\

The density matrix elements are  defined as
\begin{equation}
\densOrd{ag;bg'}{n}=\bra{ag}\hat{\rho}\ket{bg'}
\end{equation}
where the label $n$ provides the number of electron or hole excitations
needed to transform $g$ into $g'$.
Examples are $\densOrd{bg-k;ag}{1}$ and
$\densOrd{cg-k;ag+k'}{2}$. We denote the elements
as $n$-ehx elements in the following.

The von Neumann equation (\ref{EqvN}) gives
equations of motion of the type
\begin{multline}\label{EqEOMzeroth}
\imai\hbar\frac{\d}{\d t} \densOrd{bg;b'g}{0}=(E_b-E_{b'})\densOrd{bg;b'g}{0}\\
+\sum_{a,\ksa}T_{ba}(k)\densOrd{ag+k;b'g}{1}
+\sum_{c,\ksa}T_{cb}^*(k)\densOrd{cg-k;b'g}{1}\\
-\sum_{c,\ksa}\densOrd{bg;cg-k}{1}T_{cb'}(k)
-\sum_{a,\ksa}\densOrd{bg;ag+k}{1}T_{b'a}^*(k),
\end{multline}
and
\begin{multline}\label{EqEOMfirst}
\imai\hbar\frac{\d}{\d t} \densOrd{cg-\ksa;bg}{1}=
(E_c-E_b-E_k)\densOrd{cg-k;bg}{1}\\
+\sum_{b'}T_{cb'}(k)\delta_{N_k,1}\densOrd{b'g;bg}{0}
-\sum_{c'}\densOrd{cg-k;c'g-k}{0}T_{c'b}(k)\\
+\sum_{k'\sigma'\ell'}\Big[\sum_{b'}T_{cb'}(k')\densOrd{b'g-k+k';bg}{2}
+\sum_{d}T_{dc}^*(k')\densOrd{dg-k-k';bg}{2}\\
-\sum_{c'}\densOrd{cg-k;c'g-k'}{2}T_{c'b}(k')
-\sum_{a}\densOrd{cg-k;ag+k'}{2}T_{ba}^*(k')\Big].
\end{multline}
Thereby we obtain a hierarchy of $n$-ehx density matrix elements,
where the $n$-ehx elements show a phase rotation due to the energy
difference involved and are coupled to different $(n-1)$ and
$(n+1)$-ehx elements. In order to break the infinite hierarchy we
neglect all n-ehx elements with $n\ge 3$ and obtain a closed set
of equations.

For all  density matrix elements we perform a sum over all possible
lead configurations $g$ and define
\begin{equation}
\PhiOrd{b'b}{0}=\sum_g\densOrd{b'g;bg}{0},
\label{EqDefW}
\end{equation}
which are the elements of the reduced density matrix
$\rho_\mathrm{dot}$, and
\begin{equation}
\PhiOrd{ba}{1}(\ksa)=\sum_g\densOrd{bg-k;ag}{1},
\label{EqDefPhi}
\end{equation}
which describe the transitions of electrons between the leads and
the quantum dot. The particle current from the lead $\ell$ into
the structure, $J_\ell$, equals the rate of change of the
occupation in the lead. This gives
\begin{equation}
\begin{split}
J_\ell  & = -\frac{\d}{\d t}\sum_{k\sigma(\ell)}\ave{\cdag_{k}\cno_{k}}=
-\frac{\d}{\d t}\sum_{g,,b,k\sigma(\ell)}\delta_{N_k,1}\densOrd{bg,bg}{0}\\
&=-\frac{2}{\hbar}\sum_{k\sigma(\ell),cb}
\Im\left\{\sum_{g}T^*_{cb}(k)\densOrd{cg-k;bg}{1}\right\}\\
&=-\frac{2}{\hbar}\sum_{k\sigma(\ell),cb}
\Im\left\{T^*_{cb}(k)\PhiOrd{cb}{1}(k)\right\}. \label{EqCurrent}
\end{split}
\end{equation}
Thus the $\PhiOrd{ba}{1}(\ksa)$ terms describe current amplitudes.
Note that all 1-ehx terms can  be described by $\PhiOrd{ba}{1}(\ksa)$ as
$\densOrd{ag;b-kg}{1}=\left({\densOrd{bg-k;ag}{1}}\right)^*$ and
$\sum_g\densOrd{ag+k;bg}{1}=\sum_{g'}\densOrd{ag';bg'-k}{1}$.

Performing the sum $\sum_g$ to the time evolution
(\ref{EqEOMfirst}) some terms
do not directly obtain the form $\PhiOrd{ba}{n}$.
Here we approximate
\[\begin{split}
&\sum_g\delta_{N_{k},1}\rho_{b'g;bg} \approx
f_k\sum_g \rho_{b'g;bg}=f_k\PhiOrd{b'b}{0},\\
&\sum_g\densOrd{cg-k;c'g-k}{0} =\sum_{g'}\delta_{N_{k},0}
\densOrd{cg';c'g'}{0}\approx(1-f_k)\PhiOrd{cc'}{0}.
\end{split}\]
Similar approximations are done for the $\densOrd{bg-k;ag}{1}$
elements appearing in the equation for the 2-ehx terms. This
approximation is a factorization of the lead occupations, which
are assumed not to be affected by the transition processes. The
result is a closed set of differential equations for the reduced
density matrix $\PhiOrd{b'b}{0}$, the current elements
$\PhiOrd{ba}{1}(\ksa)$ and the similarly defined 2-ehx terms
$\PhiOrd{b'b}{2}(-\ksa+k'\sigma'\ell';0)$,
$\PhiOrd{ca}{2}(-\ksa-k'\sigma'\ell';0)$.

Defining a discrete set of $k$-states, one can set up a column
vector consisting of all the elements of the density-matrix
$\mathbf{\Phi}=\big(\mathbf{\Phi}^{[0]},\mathbf{\Phi}^{[1]},\mathbf{\Phi}^{[2]}\big)$,
where the sub-vectors contain all the elements of the
density-matrix with a specific $n$-value, as well as the complex
conjugates of the complex elements. The equation of motion for the
vector $\mathbf{\Phi}$ can be cast on a matrix form
\begin{equation}\label{EqMatrixForm}
\imai\hbar\frac{\d}{\d t}
\mathbf{\Phi} = \left(%
\begin{array}{ccc}
  \underline{\mathbf{E}}_{00}   & \underline{\mathbf{M}}_{01} & \underline{\mathbf{0}} \\
  \underline{\mathbf{M}}_{10}   & \underline{\mathbf{E}}_{11} & \underline{\mathbf{M}}_{12} \\
  \underline{\mathbf{0}}        & \underline{\mathbf{M}}_{21} & \underline{\mathbf{E}}_{22} \\
\end{array}%
\right)\mathbf{\Phi}=\underline{\mathbf{M}}\mathbf{\Phi}.
\end{equation}
The submatrices ${\underline{\mathbf{E}}}_{nn}$ are diagonal and
contain the energy differences between the states involved.

Now we consider the $\mathbf{\Phi}^{[2]}$ terms in a stationary
approximation, yielding
\[
\mathbf{\Phi}^{[2]} =\left(-\underline{\mathbf{E}}_{22}+\imai
0^+\right)^{-1} \underline{\mathbf{M}}_{21}\mathbf{\Phi}^{[1]},
\]
where the $\imai 0^+$ ensures causality, corresponding to the
Markov limit for the highest-order elements.

Inserting the result into Eq.~(\ref{EqMatrixForm}) leads to the matrix equation
\begin{equation}\label{EqMatrixFormRed}
\imai\hbar\frac{\d}{\d t}
\begin{pmatrix}\mathbf{\Phi}^{[0]}\\
\mathbf{\Phi}^{[1]}
\end{pmatrix}
= \begin{pmatrix}
  \underline{\mathbf{E}}_{00}   & \underline{\mathbf{M}}_{01}  \\
  \underline{\mathbf{M}}_{10}   & \underline{\mathbf{M}}_{11}
\end{pmatrix}
\begin{pmatrix}\mathbf{\Phi}^{[0]}\\
\mathbf{\Phi}^{[1]}
\end{pmatrix},
\end{equation}
where $\underline{\mathbf{M}}_{11}= \underline{\mathbf{E}}_{11} +
\underline{\mathbf{M}}_{12}
\left(-\underline{\mathbf{E}}_{22}+\imai 0^+\right)^{-1}
\underline{\mathbf{M}}_{21}$ is not diagonal. Due to the
$k$-dependence it is very laborious to express
$\mathbf{\Phi}^{[1]}$ solely in terms of $\mathbf{\Phi}^{[0]}$
thereby reducing the problem to a generalized master equation. The
explicit, and completely general, expressions for the equation of
motion for $\mathbf{\Phi}^{[0]}$ and $\mathbf{\Phi}^{[1]}$ are
given in Eqs.~(10,11) of \cite{PedersenPRB2005a} and is the main
result of that paper.\footnote{In \cite{PedersenPRB2005a},
$\PhiOrd{b'b}{0}$ and $\PhiOrd{ba}{1}(\ksa)$ are denoted
$w_{b'b}^{{}}$ and $\phi_{ba}(\ksa)$, respectively.}

The sub-matrices $\underline{\mathbf{M}}_{10}$ and
$\underline{\mathbf{M}}_{01}$ only contain elements proportional
to the tunneling amplitude $T_{ba}$, while the matrix
$\underline{\mathbf{M}}_{11}$ involves terms proportional to
$T_{ba}^2$. Thus the stationary solution of
Eq.~(\ref{EqMatrixFormRed}), together with the normalization
$\sum_b\Phi^{[0]}_{bb}=1$, will contain all powers of $T_{ba}$,
and so will the stationary occupations and coherences. That is,
the approach does not provide a systematic expansion in powers of
the tunneling coupling. For the current, which is proportional to
$T_{ba}\mathbf{\Phi}^{[1]}$, all terms up to order $T_{ba}^4$ as
well as a class of higher-order terms are taken into account.

Summarizing, the 2vN approach considers the reduced density matrix
(\ref{EqDefW}) and the current amplitudes (\ref{EqDefPhi}) as
variables, which are determined by the closed set of dynamical
equations (\ref{EqMatrixFormRed}). In deriving these, three
approximations are applied; (i) only coherent processes involving
transitions of at most two different $k$-states are considered,
i.e. all $n$-ehx terms with $n\ge 3$ are assumed to vanish, (ii)
the time-dependence of terms generating 2-ehx terms is neglected
which corresponds to the Markov limit, (iii) the level occupations
in the leads, $f_{k\sigma\ell}$, is unaffected by the couplings to
the dot, i.e. the densities in the leads and on the dot can be
factorized.

\section{Range of validity and comparison with other methods}
\label{SecValidity}
For noninteracting systems, Eq.~(\ref{EqTransmission}) gives the
correct result for arbitrary temperature and bias. In
\cite{PedersenPRB2005a} we demonstrated analytically that using
the 2vN approach, Eq.~(\ref{EqTransmission}) is fully recovered
for a single-level system. Numerically, we also found full
agreement for all ranges of parameters, including double dots
\cite{PedersenPRB2007} and the ferromagnetic Anderson model with
an applied magnetic field \cite{PedersenPreprint2008}, where
coherences are of central importance.\footnote{For practical
reasons only two-level systems have been studied so far.} Thus we
have strong indications that the 2vN method is able to treat
transport correctly for noninteracting systems over the full
temperature and bias ranges.

For interacting systems and temperatures $k_BT\gg \Gamma$ or in
the high bias limit, the 2vN reproduces the well-known results of
the methods presented in \cite{GurvitzPRB1996,NazarovPHB1993}.
This demonstrates the correct treatment of charging effects in
interacting systems.

For the Anderson model with infinite Coulomb repulsion, the 2vN
equations could be solved analytically, see
\cite{PedersenPRB2007}, and the result agree with the diagrammatic
real-time transport theory in the resonant tunneling approximation
\cite{KonigPRL1996}, where the onset of Kondo physics is observed
(see also Eq.~(2) of \cite{KonemannPRB2006}). However, in the
Kondo limit itself, the model misses the unitary limit and
unphysical results are found, as the strong correlations between
the dot and the leads are not properly reflected.

The validity of the 2vN approach for time-dependent problems has
not been carefully investigated. As the Markov approximation is
invoked, it might not be valid for strongly time-dependent
systems, where non-Markovian effects are important due to memory
effects, which are also relevant when evaluating higher-order
moments, as e.g., the noise \cite{BraggioPRL2006,FlindtPRL2008}.
As an example, the current through a single spinless level was
presented in  \cite{PedersenPRB2005a} which does not exhibit the
oscillations found from a time-dependent Green function approach
\cite{StefanucciPRB2004}.

In a recent paper, Jin {\it et al.} also consider quantum
transport in the same spirit as in the 2vN approach by keeping
correlations between the leads and the dot and performing an
expansion in the tunneling Hamiltonian \cite{JinJCP2008}. They
report a proof that they obtain the 2vN approach as a second-order
expansion.

\section{Elastic cotunneling}\label{SecElasCot}

As an example we apply the 2vN
method in the elastic cotunneling regime for a two-level spinless
system. We show that the 2vN results agree with a mean-field
solution and with a scattering result.

The system is described by the Hamiltonian
\begin{equation}\label{EqHamTL}
\begin{split}
H=&\sum_{k,\ell=L,R}E_{k\ell}c^\dag_{k\ell}c_{k\ell} +\sum_{k\ell
n} \left[V_{k\ell n}c^\dag_{k\ell}d_n+\mathrm{h.c}\right]\\
&+\sum_n E_n d^\dag_n d_n  + Ud_1^\dag d_1d_2^\dag d_2,
\end{split}
\end{equation}
with $n=1,2$ denoting the two dot states. The first term is the
Hamiltonian of the leads, and the last two terms are the two
single-particle states of the dot and the interaction between
them. The second term is the tunnel Hamiltonian with the tunneling
amplitudes $V_{k\ell n}$. Below it is assumed that $V_{k\ell
n}~=~x_{\ell n}t_{k}$, i.e. the couplings between both dot states
and the lead states have a fixed phase factor $x_{\ell n}$, and
$t_k$ is assumed to be a real number. The coupling constants are
defined as $\Gamma_{\ell n}(E)=2\pi\sum_k|V_{k\ell
n}|^2\delta(E-E_{k\ell n})=|x_{\ell n}|^2\Gamma(E)$. For the 2vN
calculations  a constant value $\Gamma$ for $|E|\leq 0.95W$ is
used, and it is  assumed that $\Gamma(E)=0$ for $|E|>W$. For
$0.95W<|E|<W$ an elliptic interpolation is applied. Using the
Hamiltonian above, the system can be in four different
many-particle states, denoted
$\ket{0},~\ket{1},\ket{2},\ket{d}=d_2^\dag d_1^\dag\ket{0}$, with
energies $0,~E_1,~E_2,~E_d$, respectively.

The transport is calculated in a setup with $E_1\ll\mu_\ell$ and
$E_2+U\gg \mu_\ell$ such that the state $\ket{1}$ is the ground
state. Sequential tunneling processes are blocked due to the
Coulomb interaction between the electrons, but a leakage current
due to elastic cotunneling processes can occur. Here, the 2vN
results are compared with a mean-field solution embedded in a
nonequilibrium Green function framework, and a scattering
formalism.

In the mean-field solution, the interaction term in the
Hamiltonian is replaced with
\begin{equation}
\begin{split}
Ud_1^\dag d_1d_2^\dag d_2\rightarrow U\Big\{&\left[d^\dag_1 d_1
\ave{d^\dag_2 d_2}+d^\dag_2 d_2 \ave{d^\dag_1
d_1}\right]\\
&-\left[d^\dag_1 d_2 \ave{d^\dag_2 d_1}+d^\dag_2 d_1 \ave{d^\dag_1
d_2}\right]\Big\},
\end{split}
\end{equation}
where the first $[\ldots]$-bracket is the Hartree term and the
second the Fock term. The occupations are calculated
self-consistently whereafter the current can be
evaluated.\footnote{The calculations are analogous to those
presented in \cite{PedersenPreprint2008}.}

For the scattering result, the elastic second-order scattering
rate, $\gamma_{11}^{RL}$, is the sum over all processes where an
electron labelled $k'L$ has been transferred from the left contact
to the state $kR$ in the right contact, leaving the dot state
unchanged. This can be realized in two different ways,
$\ket{1}\rightarrow\ket{0}\rightarrow\ket{1}$ and
$\ket{1}\rightarrow\ket{d}\rightarrow\ket{1}$, where the
amplitudes for these two processes are added coherently.

The rate is calculated as follows \cite{BruusBOOK2004}:
Assume that initially the dot is in state $\ket{1}$ and
the leads are in a state $\ket{\nu_L\nu_R}$, i.e. the initial
state is $\ket{i}=\ket{\nu_L\nu_R 1}$ with energy
$E_i=E_{\nu_L\nu_R}+E_1$. The probability for the leads to be in
the state $\nu_L\nu_R$ is denoted
$W_{\nu_L\nu_R}=W_{\nu_L}W_{\nu_R}$, as the leads are assumed
uncorrelated. In the final state an electron $k'L$ has been
transferred from left to right ending up in the state $kR$, i.e.
the final state is $\ket{f_{kk'}}=c_{kR}^\dag c_{k'L}\ket{i}$ with
energy $E_i-E_{k'L}+E_{kR}$.
According to the $T$-matrix
formalism the second-order scattering rates are
\cite{BruusBOOK2004}
\begin{equation}\label{EqGeneralRate}
\begin{split}
\gamma^{RL}_{11}=2\pi
\sum_{kk'}\sum_{\nu_L\nu_R}&W_{\nu_L\nu_R}\left|\bra{f_{kk'}}H_T\frac{1}{E_i-H_0}H_T\ket{i}
\right|^2 \\ &\times \delta(E_{f_{kk'}}-E_i),
\end{split}
\end{equation}
and after some algebra we arrive at
\begin{equation}\label{EqScatt11}
\begin{split}
\gamma^{RL}_{11} =\frac{\Gamma^2}{2\pi}\int dE
&\Big|\frac{x_{L1}^*x_{R1}}{E_1-E}+\frac{x_{R2}x_{L2}^*}{E_2+U-E}
\Big|^2\\
&\times n_F(E-\mu_L)[1-n_F(E-\mu_R)],
\end{split}
\end{equation}
where $n_F(E)=[1+e^{E/k_BT}]^{-1}$ is the Fermi function, and
energy-independent coupling constants are assumed (the Wide-Band
Limit). The rate $\gamma^{LR}_{11}$ is found by interchanging
$L\leftrightarrow R$. Below we only calculate the integral for
$k_BT=0$, but note that for finite temperatures the integral
diverges and a regularization procedure is needed
\cite{TurekPRB2002}.

For $k_BT=0$, $\mu_R=eV_\mathrm{bias}>0$ and $\mu_R=0$,  we obtain
\begin{equation}\label{EqScatt11Zero}
\gamma^{RL}_{11} =\frac{\Gamma^2}{2\pi}\int_0^{eV_\mathrm{bias}}
dE
\Big|\frac{x_{L1}^*x_{R1}}{E_1-E}+\frac{x_{R2}x_{L2}^*}{E_2+U-E}
\Big|^2,
\end{equation}
and $\gamma^{LR}_{11}=0$. Assuming the state $\ket{1}$ to be
almost completely occupied (i.e. $|E_1|/\Gamma\ll 1$ and
$(E_2+U)/\Gamma \gg 1$), the second-order elastic cotunneling
current is $I^\mathrm{el.cotun}=\frac{1}{\hbar}\gamma^{RL}_{11}$.

\begin{figure}[tb]
  \begin{center}
    \includegraphics[angle=0,width=.5\textwidth]{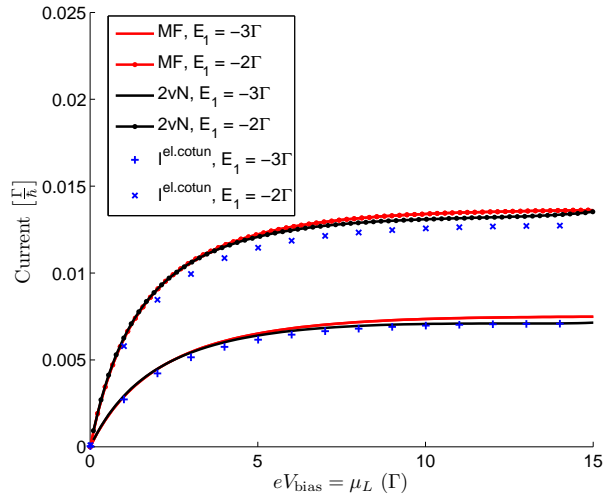}
    \caption{(Color online) The $IV$-characteristics for two different
values of $E_1$ calculated using the 2vN approach, mean-field (MF)
nonequilibrium Green functions, and, finally, the scattering
cotunneling current $I^\mathrm{el.
cotun}=\frac{1}{\hbar}(\gamma_{11}^{RL}-\gamma_{11}^{LR})$ , where
the latter is calculated for $k_BT=0$. The chemical potentials are
$\mu_L=eV_\mathrm{bias}$ and $\mu_R=0$, and the phase factors are
all equal $x_{\ell n}=1/\sqrt{2}$, i.e. $\Gamma_{\ell
n}=\Gamma/2$. The other parameters are: $E_2=16\Gamma$,
$U=20\Gamma$, $k_BT=\Gamma/10$ and $W=50\Gamma$. }
    \label{FigCotunneling}
  \end{center}
\end{figure}

In Fig.~\ref{FigCotunneling} the current versus bias voltage is
calculated in the elastic cotunneling regime using the 2vN method,
the mean-field approximation in a Green function framework, and
the second-order scattering method. For the latter, the
calculation is for vanishing temperature, while the other
calculations are for $k_BT=\Gamma/10$. For $E_1=-3\Gamma$ almost
perfect agreement between all three methods is found, while
deviations between the scattering method and the others occur for
$E_1=-2\Gamma$, which is most likely due to the fact that the
state $\ket{1}$ is not fully occupied in this case as assumed in
the derivation of the second-order scattering rate
(\ref{EqScatt11}).

In summary, from a comparison with both a mean-field solution and
a scattering formalism, we have shown that the 2vN method is able
to quantitatively describe elastic cotunneling processes even for
temperatures much lower than the energy scale set by the coupling
to leads, $k_BT\ll \Gamma$.

\section{Inelastic cotunneling}\label{SecInelasCot}

\begin{figure}[tb]
  \begin{center}
    \includegraphics[angle=0,width=.5\textwidth]{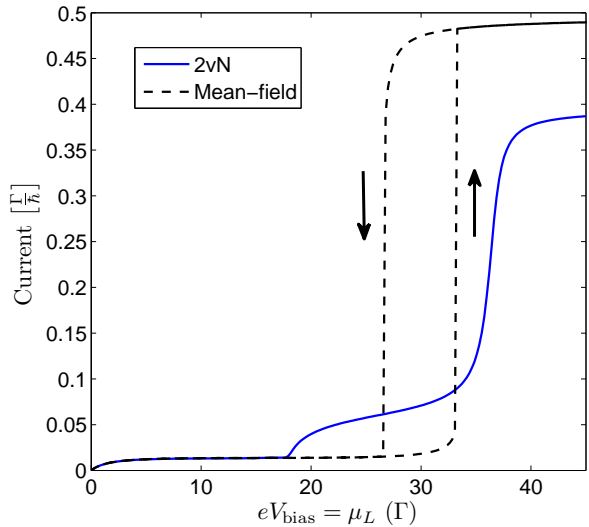}
    \caption{(Color online) The $IV$-characteristic over the full bias range for $E_1=-2\Gamma$ calculated
    within the mean-field approximation and using the 2vN approach. The
    other parameters are as in Fig.~\ref{FigCotunneling}. The arrows indicate the direction of the bias sweep
    in the mean-field calculation.}
    \label{FigBistable}
  \end{center}
\end{figure}

For higher bias voltages the agreement between the 2vN approach
and the mean-field solution is assumed to be less perfect, as the
latter can lead to bistable solutions \cite{HorvathPRB2008}, which
are not present in a generalized master equation approach.
Fig.~\ref{FigBistable} shows a comparison between the 2vN approach
and the mean-field solution over the full bias range for
$E_1=-2\Gamma$, with the rest of the parameters as in
Fig.~\ref{FigCotunneling}.

Considering first the 2vN result, the curve shows increased
current when the bias matches the energy difference between the
levels, $eV_\mathrm{bias}=|E_1-E_2|=18\Gamma$. This is due to the
onset of inelastic cotunneling \cite{BruusBOOK2004}, which leads
to a population of the excited state, $\ket{2}$. After the
inelastic cotunneling process,  additional cotunneling-assisted
sequential tunneling processes can occur
\cite{GolovachPRB2004,SchleserPRL2005}. Finally, at
$eV_\mathrm{bias}=E_2+U=36\Gamma$ sequential tunneling through the
upper level becomes possible. The value for the current is
consistent with the master equation result ($3\Gamma/8\hbar$) plus
additional cotunneling through the lower level.

In contrast, the mean-field solution misses the onset of inelastic
cotunneling and instead a bistable behaviour is observed, and it
also overestimates the current after the onset of sequential
tunneling. Both is due to an insufficient description of the
Coulomb interaction and emphasizes the need for a method which can
describe both higher-order tunneling processes (as elastic and
inelastic cotunneling) and many-particle interactions.

\section{Summary}

The second order von Neumann approach provides a quantitative
description of transport through nanostructures for arbitrary bias
and temperatures above the Kondo temperature. In the low-bias
regime, the elastic cotunneling current is in very good agreement
with both the second-order scattering result and the mean-field
solution, even for temperatures much lower than the energy scale
set by the coupling to leads. Inelastic cotunneling is also well
captured by the 2vN approach, whereas a mean-field solution within
a nonequilibrium Green functions framework shows an artificial
bistability in this regime.


\end{document}